\documentclass{article}
\usepackage{spconf,amsmath,epsfig}
\DeclareGraphicsExtensions{.png, .pdf}
\usepackage{graphicx, dblfloatfix}
\usepackage{url}
\usepackage{subcaption}
\usepackage{graphicx}
\usepackage{hyperref}
\usepackage{amssymb}
\usepackage{balance}
\usepackage[none]{hyphenat}

\newcommand{\squeezeup}{\vspace{-2.5mm}}
\newcommand{\squeezeupf}{\vspace{-1.5mm}}

\let\OLDthebibliography\thebibliography
\renewcommand\thebibliography[1]{
  \OLDthebibliography{#1}
  \setlength{\parskip}{0pt}
  \setlength{\itemsep}{0pt plus 0.3ex}
}

\begin{document}\sloppy
\ninept

\def\x{{\mathbf x}}
\def\L{{\cal L}}

\title{Perceptual evaluation on Audio-visual dataset of 360 content
}
%
\name{Randy F Fela\textsuperscript{1,4}, Andr\'eas Pastor\textsuperscript{2}, Patrick Le Callet\textsuperscript{2}, Nick Zacharov\textsuperscript{3}\sthanks{The work was performed whilst the author was at  FORCE Technology - SenseLab. Currently the author is at Meta Reality Labs.}, Toinon Vigier\textsuperscript{2}, Søren Forchhammer\textsuperscript{4}}

\address{\textsuperscript{1}SenseLab, FORCE Technology, 2970 Hørsholm, Denmark \\\textsuperscript{2}Nantes Université, Ecole Centrale Nantes, CNRS, LS2N, UMR 6004, F-44000 Nantes, France\\\textsuperscript{3}Meta Reality Labs -- Meta, Paris, France\\\textsuperscript{4}DTU Electro, Technical University of Denmark, 2800 Kgs. Lyngby, Denmark}

\maketitle

\begin{abstract}
To open up new possibilities to assess the multimodal perceptual quality of omnidirectional media formats, we proposed a novel open source 360 audiovisual (AV) quality dataset. The dataset consists of high-quality 360 video clips in equirectangular (ERP) format and higher-order ambisonic (4\textsuperscript{th} order) along with the subjective scores. 
Three subjective quality experiments were conducted for audio, video, and AV with the procedures detailed in this paper. Using the data from subjective tests, we demonstrated that this dataset can be used to quantify perceived audio, video, and audiovisual quality. The diversity and discriminability of subjective scores were also analyzed. Finally, we investigated how our dataset correlates with various objective quality metrics of audio and video. Evidence from the results of this study implies that the proposed dataset can benefit future studies on multimodal quality evaluation of 360 content.

\end{abstract}

\section{Introduction}

\label{sec:intro}
Omnidirectional media formats (as 360 video with spatial audio) offers a more immersive experience than traditional audiovisual presentations, leading to an increasing level of adoption across multimedia service platforms \cite{choi2017information}. To achieve a high-level user experience for multimedia services, perceived quality needs to be better understood, which is commonly evaluated through computational objective metrics validated by a series of subjective experiments. While multisensory evaluation is necessary to perform a higher degree of model prediction, a study in this integral quality is relatively unexplored due to the lack of multimodal dataset.

Public datasets of 360 video that contains subjective quality scores are IVQAD \cite{duan2017ivqad}, VR-VQA \cite{xu2018assessing}, and VQA-ODV \cite{li2018bridge}. However, these datasets are limited partially due to the 4K video resolution \cite{duan2017ivqad} and/or sourced from streaming services with unknown quality control \cite{xu2018assessing, li2018bridge}. Meanwhile, large spatial audio datasets can be found in projects of 3D-MARCo \cite{lee2019open}, EigenScape \cite{green2017eigenscape}, and ARTE \cite{weisser2019ambisonic}. However, all mentioned datasets were mainly focused on a single modality and none of the spatial audio datasets performed audio quality evaluation.

\begin{figure}[!t]
\centering
    \includegraphics[width=0.45\textwidth]{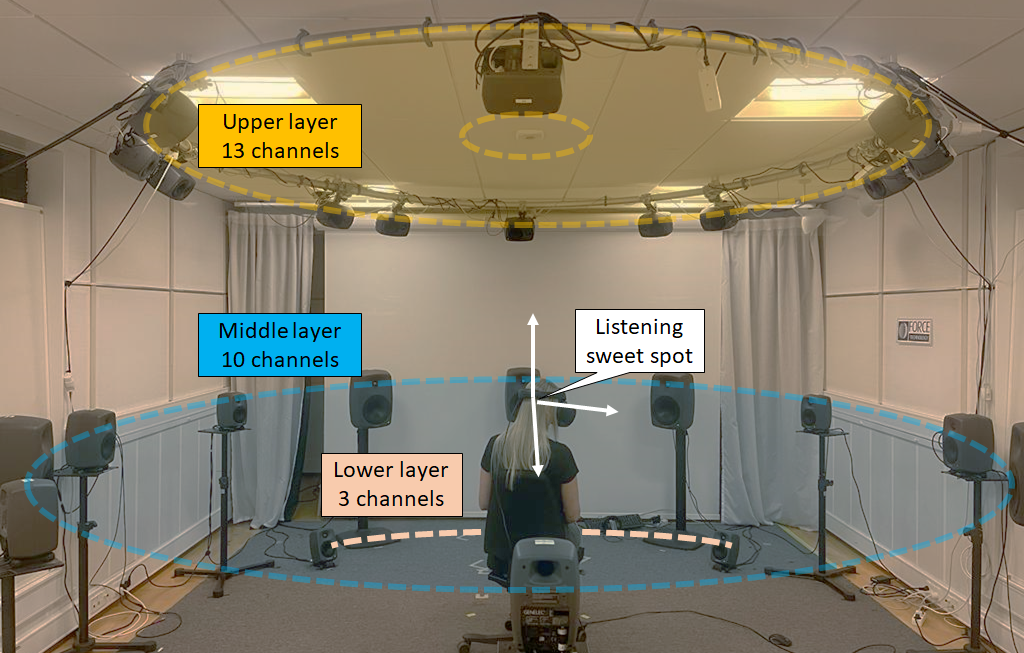}
\squeezeup
\caption{Experimental setup for subjective experiments.}
\squeezeup
\label{fig:1}
\end{figure}

To better understand the quality of audio-visual multimodal quality, a high quality multimodal dataset is required. For traditional multimedia applications such as 2D video and channel-based audio, several audiovisual datasets exist such as PLYM \cite{goudarzi2010audiovisual}, TUM 1080p50 \cite{keimel2012tum}, VQEG \cite{pinson2012influence}, Vienna Made for Mobile \cite{robitza2012made}, VTT \cite{maki2013reduced}, and INRS \cite{demirbilek2016inrs}. 
In comparison to traditional media formats, it is important to investigate immersive multimedia formats as they provide spatial information, allowing users an increased spatial experience. Several databases have been proposed and studied for different purposes such as attribute evaluation \cite{rummukainen2014categorization}, developing media player \cite{bailer2015multi}, audio generation \cite{rana2019towards}, and audiovisual attention \cite{chaoicmew2020}. However, there is a 
current shortage of high quality immersive audiovisual datasets and the absence of subjective quality scores.

\begin{figure*}[!ht]
\centering
 \mbox{ \parbox{1\textwidth}{
  \begin{minipage}[b]{0.16\textwidth}
  {\label{fig:dmos_corr_img}\includegraphics[width=\textwidth]{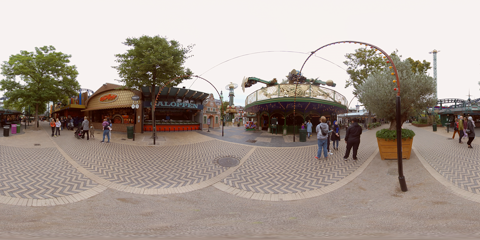}}
  \end{minipage}
  \begin{minipage}[b]{0.16\textwidth}
  {\label{fig:dmos_corr_img}\includegraphics[width=\textwidth]{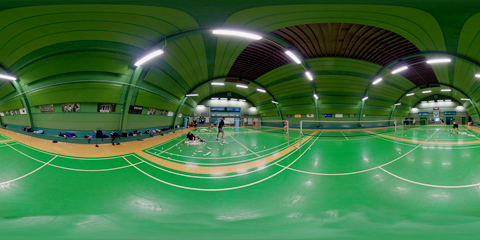}}
  \end{minipage}
  \begin{minipage}[b]{0.16\textwidth}
  {\label{fig:dmos_corr_img}\includegraphics[width=\textwidth]{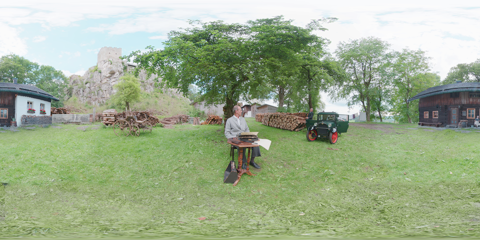}}
  \end{minipage}
  \begin{minipage}[b]{0.16\textwidth}
  {\label{fig:dmos_corr_img}\includegraphics[width=\textwidth]{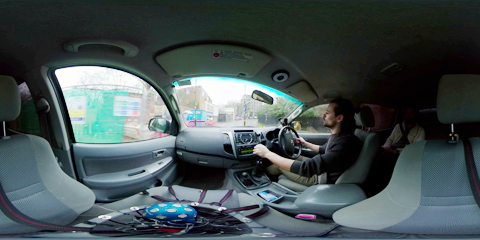}}
  \end{minipage}
  \begin{minipage}[b]{0.16\textwidth}
  {\label{fig:dmos_corr_img}\includegraphics[width=\textwidth]{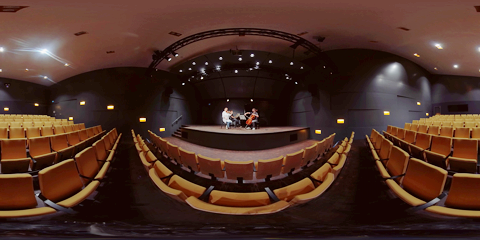}}
  \end{minipage}
  \begin{minipage}[b]{0.16\textwidth}
  {\label{fig:dmos_corr_img}\includegraphics[width=\textwidth]{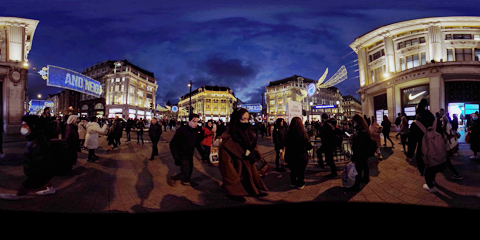}}
  \end{minipage}\\
  \begin{minipage}[b]{0.16\textwidth}
  {\label{fig:dmos_corr_img}\includegraphics[width=\textwidth]{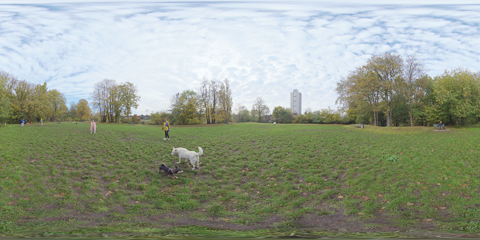}}
  \end{minipage}
  \begin{minipage}[b]{0.16\textwidth}
  {\label{fig:dmos_corr_img}\includegraphics[width=\textwidth]{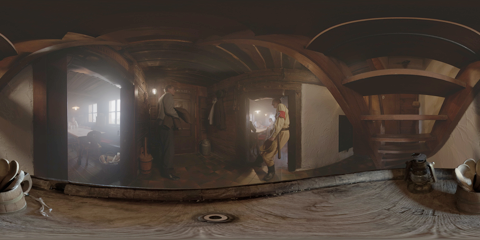}}
  \end{minipage}
  \begin{minipage}[b]{0.16\textwidth}
  {\label{fig:dmos_corr_img}\includegraphics[width=\textwidth]{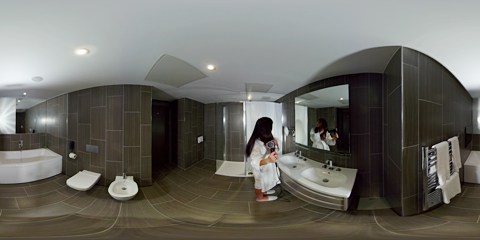}}
  \end{minipage}
  \begin{minipage}[b]{0.16\textwidth}
  {\label{fig:dmos_corr_img}\includegraphics[width=\textwidth]{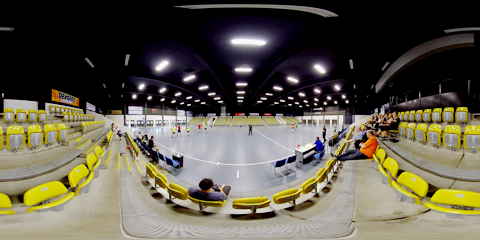}}
  \end{minipage}
  \begin{minipage}[b]{0.16\textwidth}
  {\label{fig:dmos_corr_img}\includegraphics[width=\textwidth]{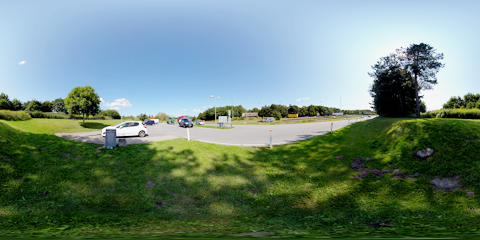}}
  \end{minipage}
  \begin{minipage}[b]{0.16\textwidth}
  {\label{fig:dmos_corr_img}\includegraphics[width=\textwidth]{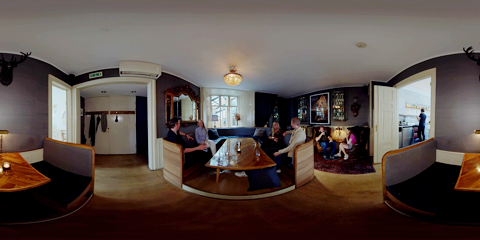}}
  \end{minipage}\\
  \begin{minipage}[b]{0.16\textwidth}
  {\label{fig:dmos_corr_img}\includegraphics[width=\textwidth]{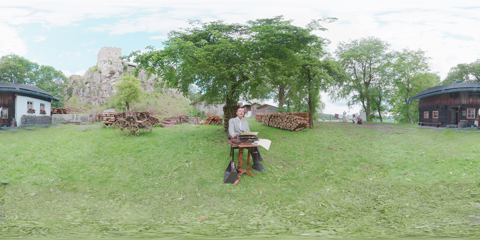}}
  \end{minipage}
  \begin{minipage}[b]{0.16\textwidth}
  {\label{fig:dmos_corr_img}\includegraphics[width=\textwidth]{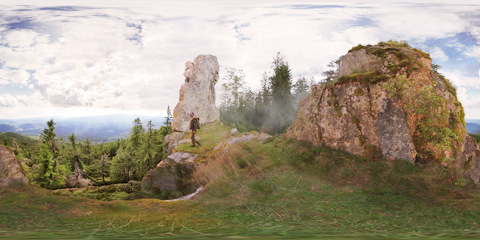}}
  \end{minipage}
  \begin{minipage}[b]{0.16\textwidth}
  {\label{fig:dmos_corr_img}\includegraphics[width=\textwidth]{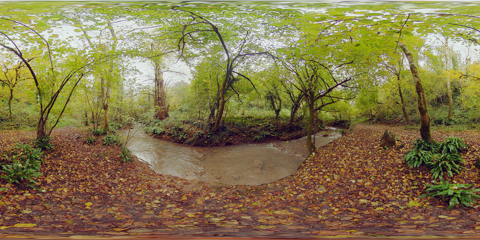}}
  \end{minipage}
  \begin{minipage}[b]{0.16\textwidth}
  {\label{fig:dmos_corr_img}\includegraphics[width=\textwidth]{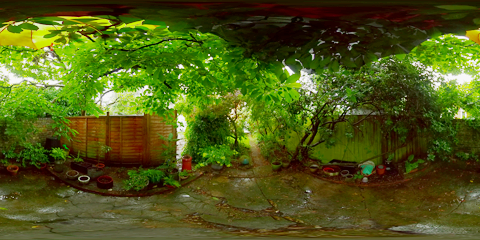}}
  \end{minipage}	
   }}
\squeezeup
\caption{Equirectangular video previews of the proposed dataset.}
\squeezeupf
\squeezeup

\label{fig:2}
\end{figure*}

In this work, we present a dataset which consists of recorded 8K 360 video with higher/4\textsuperscript{th} order ambisonic (HOA) audio along with mean opinion scores collected from audio, video, and audiovisual subjective experiments. The focus of this paper is to present our investigation on subjective scores which include overall mean-CI analysis, SOS analysis benchmarked with existing datasets, and evolution of accuracy performance. 
After the investigation of subjective scores, a number of audio and video objective quality metrics were computed to investigate how these metrics perform in correlation to our subjective data, and to identify potential metrics towards the development of audiovisual quality metrics. To the best of our knowledge, this is the first recorded audiovisual dataset created to support perceptual quality research in immersive audiovisual content. 

The remainder of this paper is organized as follows. Section 2 includes the description of the database, encoding and decoding steps to create processed sequences, and subjective evaluation procedures. We discuss our findings from subjective evaluations and objective quality metrics in Section 3. Finally, the conclusion is addressed in Section 4.


\section{Materials and Evaluations}
\squeezeupf
In this section, we describe HOA--SSR database, encoding parameters to create processed audio/video sequences, and evaluation of objective quality metrics and subjective data.

\squeezeup
\subsection{360 audiovisual stimuli}
This audiovisual (AV) dataset consists of 12 recorded audiovisual scenes selected from HOA--SSR database which contains immersive AV contents with unique characteristics i.e. nature--mechanical, indoor--outdoor, static--dynamic, traffic--quiet, impulsive--steady, and speech--music. The video scenes were recorded using an Insta360 Pro2, a professional spherical 360 camera, that consists of 6 camera lenses to capture every angle of a scene at once. The final video format for the dataset was provided in .mov container with 8K resolution (7680x3840), 30fps, 8-bit color depth, and in YUV422. The audio signals were recorded using the em32 Eigenmike, a spherical microphone with an array of 32 microphones. The output of these recordings was in a raw 32-channel ambisonic A-format then processed in 4\textsuperscript{th} order ambisonic B-format AmbiX (25 channels) in ACN ordering and SN3D normalization. All audio files were in PCM 1152 kbps/channel, 24bit and 48kHz. In terms of spatial characteristics of a microphone, a previous study reported the highest directional accuracy of em32 compared to all other high order sound field microphones \cite{bates2016comparing, bates2017comparing}. The AV dataset used in this study as illustrated in Figure \ref{fig:2} is publicly available upon request\footnote{\url{https://bit.ly/HOA-SSR-Dataset}}. 

In order to cover a use case with cinematic VR video, additional 4 AV stimuli were provided from joint work of Vtopia360\footnote{\url{https://vtopia360.com/}} and VRTonung\footnote{\url{https://www.vrtonung.de/en/}}. All videos were in the same quality and format as the HOA--SSR and the audio signals were recorded by using an ORTF-3D microphone\footnote{\url{https://schoeps.de/en/products/surround-3d/ortf-3d.html}} mixed into 24 channel NHK layouts and provided in ADM format. While in principle, the ORTF-3D provided superior localization accuracy \cite{guastavino2007spatial}, it was only compared to first order ambisonic format. Higher-order ambisonic format will increase the spatial resolution, hence improving localization accuracy. In our study, the use of these two types of recording techniques was considered equivalent since only internal quality (e.g. bitrate) was evaluated without any comparison of recording technique and assessment of attribute quality.  

\squeezeup

\subsection{Stimulus preparation: Encoding and decoding}

From the original raw format YUV422, all video sources were downscaled to a resolution of 6144x3072 and in YUV420 format due to the maximum limit of our playback system. We used libx265 (H.265/HEVC) in FFmpeg 4.4\footnote{\url{https://github.com/GyanD/codexffmpeg/releases/tag/4.4}} to encode the source videos into three resolutions i.e. 6K (6144x3072), 4K (3840x1920), 2K (1920x1080) and 4 QPs (0, 22, 28, 34), resulting in 12 encoding parameters and 192 video stimuli in total. Meanwhile, the ambisonic audio sources were encoded from 32-channel A-format into 25-channel 4\textsuperscript{th} order ambisonic in ambiX. All audio was encoded using FFmpeg with AAC-LC encoder into four different bitrates/channels (16kbps, 32kbps, 64kbps, PCM/reference) resulting in 64 audio stimuli in total. Due to the limitation of the channel number in the AAC encoder, the audio channels were split into six groups of 4-channel and 1 mono channel prior to encoding and re-grouped thereafter. Ambisonic audio files were decoded by using the All-Round Ambisonic Decoding (AllRAD) algorithm as proposed in \cite{zotter2012all} into 26 multichannel loudspeaker setups that follow the standard in \cite{ITURBS2159}. AllRAD provides energy preservation across direction and average localization sharpness. Only decoding part was required for NHK audio format. For the experiments, 20s length out of 1 minute original duration was selected based on spectral frequency profile. 

\begin{figure}[!t]
\centering
\includegraphics[width=\linewidth]{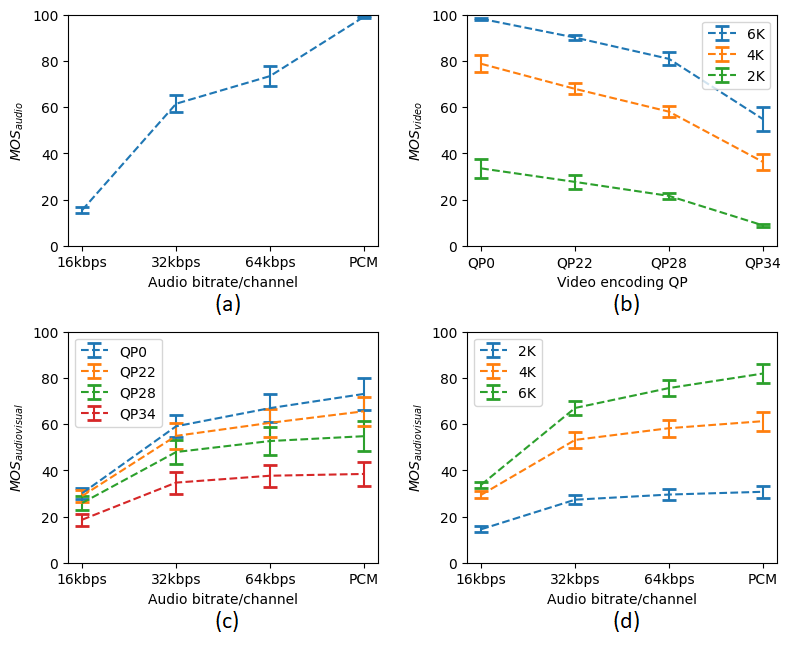}
\squeezeupf
\squeezeupf
\squeezeupf
\squeezeup
\caption{Mean opinion score (CI 95\%) of (a) audio, (b) video, and (c-d) audiovisual (AV).}
\squeezeup
\squeezeup
\label{fig:mean_mos_analysis}
\end{figure}

\begin{figure*}[]
\centering
 \mbox{ \parbox{1\textwidth}{
  \begin{minipage}[b]{0.33\textwidth}
  {\label{fig:SOS_audio}\includegraphics[width=\textwidth]{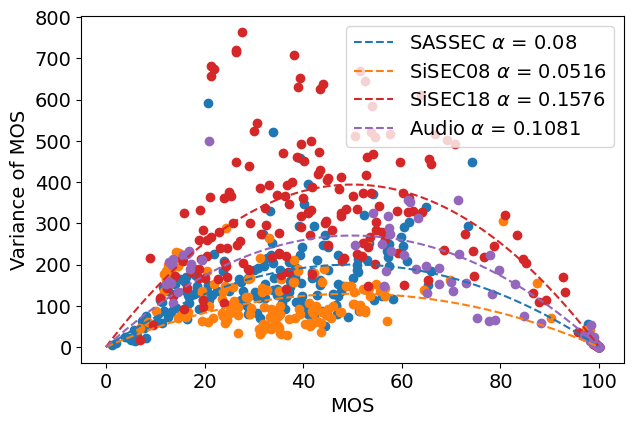}}
  \end{minipage}
  \hfill
  \begin{minipage}[b]{0.33\textwidth}
  {\label{fig:SOS_video}\includegraphics[width=\textwidth]{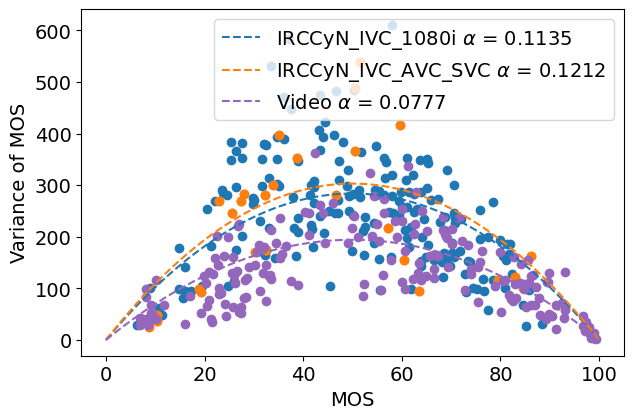}}
  \end{minipage}
  \hfill
  \begin{minipage}[b]{0.33\textwidth}
  {\label{fig:SOS_audiovisual}\includegraphics[width=\textwidth]{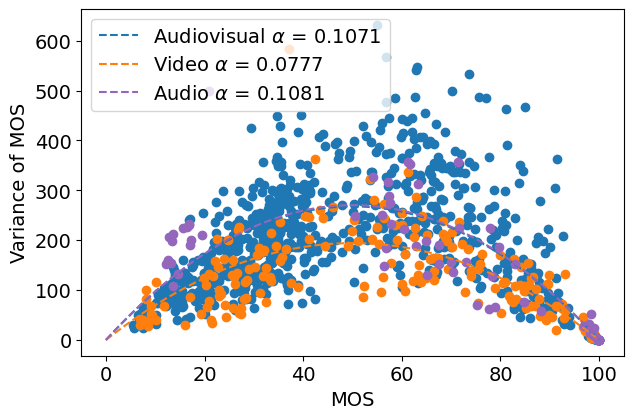}}
  \end{minipage}
  }}
\squeezeup
\caption{SOS analysis for audio (left), video (middle) and audiovisual (right) experiments against existing datasets of the literature.}
\squeezeup

\label{fig:4}
\end{figure*}

\squeezeup
\subsection{Subjective evaluations}
Twenty-one selected and trained assessors (13 males, 8 females) with age range 22 -- 37 years (mean = 27.9, std = 4.0) were invited to participate in three consecutive subjective experiments carried out at FORCE Technology SenseLab including I: listening/audio, II: viewing/video, and III: AV test. 
The experiments were performed in a standardized listening room that meets the acoustical requirements of EBU 3276 \cite{EBU3276} and ITU-R BS.1116-3 \cite{ITURBS1116-3}, compliants for listening and VR experiment with head-mounted display. The experimental setup for AV experiment is depicted in Fig \ref{fig:1}.

To avoid bias occurring between auditory and visual memory, the subjective test ordering was audio, then video, and finally AV experiment. We used SenseLabOnline 4.2 \cite{le2013senselabonline} as the user interface and ran double blinded-randomized trials. The participant was seated on a rotating chair, located in the acoustically sweet spot, and used a pad controller to perform the test. The user interface was displayed on a projection screen for experiment I and virtually in the Head Mounted Display (HMD) for experiments 2 and 3. Multiple stimulus with hidden reference without anchor (modified MUSHRA) was used to generalize the common evaluation methodology found in SAMVIQ \cite{kozamernik2005samviq} for video and MUSHRA \cite{ITURBS1534-3} for intermediate audio quality. Each assessor was asked to rate their overall perceived quality of the audio, video, and audiovisual to a continuous rating scale between 0 and 100 categorized into 5 labels (Bad, Poor, Fair, Good, Excellent). We limited the number of stimuli on each trial set to 7 
according to Miller's Law \cite{miller1956magical}.

In experiment I, the participants rated the audio quality of sound stimuli reproduced over a 26-channel setup of 8040A Genelec loudspeakers. The sound level of stimuli were subjectively calibrated to 65 -- 73 dB for most comfortable loudness, depending on the samples. Experiment II was a viewing test, where the task was to assess the perceived quality of 360 videos. Visual stimuli were displayed in VR using a Samsung Odyssey+ HMD which has a display resolution of $1440\times1600$ per eye, 110$^\circ$ horizontal field of view and 90Hz refresh rate. Finally, experiment III was an AV test where the participants rated the integrated AV quality in overall experience. One subject was withdrawn in the middle of experiment II due to a comfort issue, therefore the total number of participants was 20 subjects (12 males, 8 females; mean = 28.1, std = 4.0) for the last two experiments.

\squeezeup
\subsection{Evaluation of objective quality metrics}

We evaluated 360 video quality metrics: S-PSNR \cite{yu2015framework}, WS-PSNR \cite{sun2017weighted}, and CPP-PSNR \cite{zakharchenko2016quality}. In addition, we calculated VMAF \cite{vmaf, vmaf2017rassool} and its components, VIF and DLM \cite{vif, dlm} as shown in previous studies \cite{orduna2019video, fela2021perceptual}. Lastly, we calculated 2D image quality metrics: PSNR, SSIM\cite{ssim}, and MS-SSIM\cite{wang2003multiscale}. 
For audio quality metrics, we evaluated PEAQ \cite{ITURBS1387,thiede2000peaq}, ViSQOL \cite{hines2015visqolaudio, sloan2017objective}, and AMBIQUAL \cite{narbutt2020ambiqual} for W-channel component of ambisonic as it represents both ambient and direct signals, and center channel of NHK format. Only listening quality feature of AMBIQUAL was computed. Three frequency bands were calculated for ViSQOL: denoted as ViSQOL\textsubscript{nb} (150 -- 3400Hz), ViSQOL\textsubscript{wb} (50 -- 8000Hz), and ViSQOL\textsubscript{aswb} (50 -- 16000Hz).

\section{Results and Analysis}
\squeezeup

\subsection{Perceived audio, video, and audiovisual quality}

Figure \ref{fig:mean_mos_analysis} depicts the results obtained from audio, video, and AV subjective experiments to analyze perceptual difference between encoding parameters. The data is presented as Mean Opinion Score (MOS) over all stimuli with 95\% confidence interval (CI). In listening test (Figure \ref{fig:mean_mos_analysis}(a)), there is a large perceptual gap between 16 and 32 kbps, and between 64kbps and PCM, where PCM is the reference. Although the difference is statistically significant, mean score difference is perceptually small between 32 and 64kbps. Video quality test presented in Figure \ref{fig:mean_mos_analysis}(b) shows by nature of perceived video quality over encoding parameters, that MOS score is, as expected, higher in lower QP and higher resolution, and vice versa. It can also be seen that there is a significant difference between QPs at each resolution. 
Finally, Figure \ref{fig:mean_mos_analysis}(c-d) show the results from the audiovisual experiment averaged over resolutions and QPs in relation to audio bitrates, respectively. It is shown that perceived audiovisual quality difference in all bitrates is statistically different if video quality is higher (QP $\leq$ 28, resolution $\geq$ 4K).
\squeezeup

\subsection{SOS analysis}

When investigating perceptual quality scores, one may gain the information from the collected data by observing user rating diversity. Therefore, we employed the Standard deviation of Opinion Scores (SOS) hypothesis, which postulates a quadratic relationship between the MOS and SOS${^2}$ which depends only on one parameter $\alpha$. We modified the equation formulated in \cite{hossfeld2011sos} for Absolute Category Rating (ACR) use case 1 -- 5 to our continuous rating scale 0 -- 100,
\squeezeup

\begin{equation}
    \sigma^{2}(MOS) = \alpha(MOS - 0)(100 - MOS)
\end{equation}

Several existing audio and video datasets were benchmarked with ours to investigate rating diversity in typical studies. We selected three datasets available in the SEBASS-DB database named SASSEC, SiSEC08 and SiSEC18 datasets originally collected for evaluation of audio source separation algorithms \cite{kastner2019efficient}, and IRCCyN datasets for video quality evaluation case \cite{pechard2008suitable}. We focused on the datasets which use multiple stimulus methodologies, MUSHRA for audio and SAMVIQ for video. However, no available AV dataset exists with this rating paradigm.

Figure \ref{fig:4} summarizes SOS analysis for audio and video experiments benchmarked with existing datasets, and Figure \ref{fig:4}(right) represents SOS analysis for our three experiments. Our observation is on the SOS parameter $\alpha$. In audio experiments, $\alpha$ is 0.051 for SiSEC08 and 0.197 for SiSEC18. 
In comparison, our audio dataset has an $\alpha$ of 0.1081, which is in the range observed for audio studies with MUSHRA. The SOS score and the $\alpha$ value indicate that the dataset has consistent rating scores with low diversity.

For video datasets, $\alpha$ is in the same value range as both IRCCyN datasets, since $\alpha$ for our dataset is 0.077. Compared to our dataset, the diversity of user judgments in benchmark datasets is higher for scores between 20 and 60. By plotting our three datasets in Figure \ref{fig:4}(right), even if the number of systems tested and signals rated are different, it is shown that the $\alpha$ values are small within the range of 0.077 and 0.108. These small values of $\alpha$ indicate the consistency of user experience quality, as we argue also as the benefit of using trained assessors \cite[p. 105]{bech2007perceptual}.

Overall, the difference for all tested datasets in this study is considerably low compared to previous audio studies, which ranges between 0.269 and 0.590 \cite{kastner2019efficient}. A wide range of benchmark studies also showed that the range of $\alpha$ in image and video QoE is 0.037 -- 0.590. Nevertheless, the primary purpose of the SOS analysis in this study is to support comparisons and reliability checks between subjective studies. A large and in-depth benchmarking study is required in order to categorize the level of diversity based on $\alpha$ and SOS score, particularly in multiple stimulus rating methodologies.

\begin{figure}[]
\centering
\includegraphics[width=0.8\linewidth]{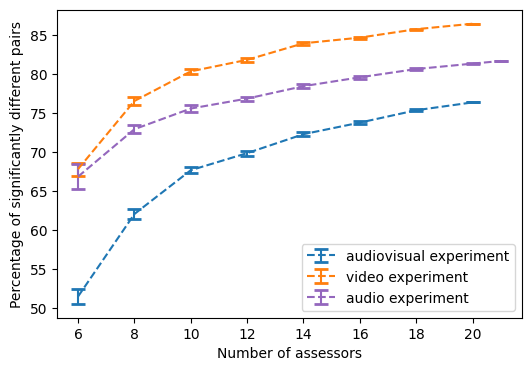}
\squeezeup
\caption{The evolution of the percentage of significantly different pairs (y-axis) with an increasing number of assessors (x-axis) for the 3 experiments: audio, video, and AV. Over 100 simulations, the curves represent mean percentages and the error bars represent 95\% confidence intervals.}

\squeezeup
\squeezeup
\label{fig:5}
\end{figure}
\squeezeup

\subsection{Subjective scores discriminability analysis}
As suggested in \cite{nehme2020comparison}, we can examine the evolution of discriminability for subjective scores with an increasing number of assessors. A two-sample Wilcoxon test is performed on all the possibles pairs of stimuli and a $p_{value}$ of 0.05 is used to compute the percentage of pairs significantly different. The number of possible pairs for audio, video, and AV experiments is 2016 pairs, 18336 pairs, and 294528 pairs, respectively. The result is presented in Figure \ref{fig:5} with 95\% confidence interval over 100 simulations.

Regarding the overall trend of the curves, video and AV experiments show the highest and lowest discriminability, respectively. Audio and video tests started with similar discriminability, where audio CI is larger than video. However, by increasing the number of assessors, the rate of discriminability of video is higher than audio, which is shown by the gap between the two curves. 
This could be that the stimuli and step sizes between video stimuli are larger and easier for assessor compared to audio. For AV task, two modalities were used, making a larger cognitive load, thus the task of evaluating critically both AV was harder, hence the results in Figure 5. As previously investigated by \cite{fela2020towards}, perceptual evaluation of single modality is less complex compared to multiple modalities. 

\squeezeup
\subsection{Objective quality metrics}
The correlation between objective metrics and subjective data is presented in Table \ref{tab:1} and \ref{tab:2}, respectively, for audio and video quality metrics. We computed Pearson, Spearman and Kendall correlation coefficients. In addition, we ran statistical pairwise analysis on the performances of the different metrics, using the indicators presented in \cite{krasula2016accuracy}: percentage of correct classification, C\textsubscript{0}\%, from pairs with statistical significance differences, and AUCs from ROC analysis (AUC\textsubscript{BW} and AUC\textsubscript{DS}). 
In the tables, underlined metrics are not significantly different compared to the best performing metric reported in bold.


From table of audio quality metrics, the best performing metric on the audio dataset is ViSQOL\textsubscript{aswb}, where we use the largest frequency bands in the calculation.
PEAQ has the lowest score followed by AMBIQUAL which competing with ViSQOL\textsubscript{wb}. It is well known that AMBIQUAL was specifically designed to compute listening quality and localization accuracy of ambisonic signal as the performance proved in \cite{narbutt2020ambiqual} for low bitrate codec in 1\textsuperscript{st} and 3\textsuperscript{rd} order ambisonic. However, the performance of AMBIQUAL compared to other metrics, especially its predecessor (ViSQOL) remains unexplored.

For video quality metrics, VMAF 4K, and the bootstrapped version (VMAF B) outperforms other metrics. It is also interesting to see the performance of VMAF Neg on part with the previous one. This finding is supported by other studies that demonstrated the superiority of VMAF 4K in terms of perceptual correlation in 360 video compared to other video quality metrics \cite{orduna2019video, fela2021perceptual}.
Metrics based on PSNR and directly optimized for 360 contents (S-PSNR, WS-PSNR, CPP-PSNR) are not providing any gain compared to their 2D counterparts computed on the ERP.

\begin{table}[]

\caption{Performances of Full-Reference metrics to relation to the DMOS scores of the audio dataset. Bold best performance score and underlined scores are not significantly different from the best score.}

\scriptsize
\centering
\begin{tabular}{|c|c|c|c|c|c|c|} 
\hline
Metric & PLCC & SRCC & KRCC & C$_0\%$ & AUC$_{BW}$ & AUC$_{DS}$ \\
\hline
Ambiqual & 0.878 & 0.905 & 0.746 & 0.933 & 0.989 & \underline{0.916} \\
PEAQ & 0.753 & 0.753 & 0.552 & 0.815 & 0.944 & 0.851 \\
ViSQOL$_{nb}$ & 0.864 & 0.884 & 0.720 & 0.914 & 0.970 & 0.851 \\
ViSQOL$_{wb}$ & 0.897 & 0.912 & 0.755 & 0.938 & 0.981 & \underline{0.885} \\
ViSQOL$_{aswb}$ & \textbf{0.924} & \textbf{0.938} & \textbf{0.800} &  \textbf{0.958} & \textbf{0.995} & \underline{\textbf{0.921}} \\
 \hline
\end{tabular}

\label{tab:1}
\end{table}

\begin{table}[]
\scriptsize
\caption{Performances of Full-Reference metrics to relation to the DMOS scores of the video dataset. Bold best performance score and underlined scores are not significantly different from the best score.}

\centering
\begin{tabular}{|c|c|c|c|c|c|c|} 
\hline
Metric & PLCC & SRCC & KRCC & C$_0\%$ & AUC$_{BW}$ & AUC$_{DS}$ \\
\hline
VMAF HD & 0.859 & 0.927 & 0.767 & 0.924 & 0.973 & 0.760 \\
VMAF 4K & \underline{\textbf{0.919}} & \underline{\textbf{0.957}} & \underline{\textbf{0.822}} & \underline{\textbf{0.954}} & \underline{0.983} & \underline{0.828} \\
VMAF B & \underline{0.915} & \underline{0.955} & \underline{0.816} & \underline{0.952} & \underline{0.988} & \underline{0.824} \\
VMAF Neg & \underline{0.917} & \underline{0.957} & \underline{0.819} & \underline{0.954} & \underline{\textbf{0.989}} & \underline{\textbf{0.830}} \\
DLM & 0.893 & 0.938 & 0.787 & 0.935 & 0.980 & 0.788 \\
$VIF_{scale0}$ & 0.770 & 0.765 & 0.586 & 0.826 & 0.910 & 0.691 \\
SSIM & 0.693 & 0.823 & 0.645 & 0.856 & 0.922 & 0.680 \\
MS-SSIM  & 0.671 & 0.843 & 0.662 & 0.867 & 0.921 & 0.648\\
PSNR & 0.616 & 0.719 & 0.538 & 0.800 & 0.891 & 0.678 \\
\hline
S-PSNR  & 0.628 & 0.743 & 0.559 & 0.811 & 0.902 & 0.691 \\
WS-PSNR & 0.617 & 0.720 & 0.538 & 0.800 & 0.892 &  0.682 \\
CPP-PSNR & 0.622 & 0.731 & 0.547 & 0.805 & 0.897 & 0.686 \\
 \hline
\end{tabular}

\squeezeup
\squeezeup
\label{tab:2}
\end{table}

\squeezeup
\section{Conclusion}
\squeezeup
In this paper we present an audiovisual dataset comprising 360 video and ambisonic spatial audio with associated subjective scores. The work focused on the exploratory analysis of subjective data to understand 1) the overall perceptual difference between each encoding parameter as perceived by assessors, 2) the span of subjective scores, and 3) the improvement of subjective scores accuracy as function of assessor number. Furthermore, we showed the performance of the dataset in relation to a set of objective quality metrics for audio and video.

The findings provided in this research confirm that there are perceptual differences for different encoding parameters. 
The SOS analysis confirms that our dataset has a low $\alpha$ value and variance for stimuli in the middle of the rating scale, proving the quality of the proposed dataset. We also found that the $\alpha$ value for audio part of the dataset is comparable to other works that used MUSHRA methodology. However, the threshold remain unclear for $\alpha$ value in this application and further benchmark analysis is required. In subjective scores discriminability analysis, video experiment was placed the highest, followed by audio and AV experiment. All curves have a low CI, with a steady trend after 12 number of assessors, confirming our choice of 20 assessors on each task. Lastly, objective metrics analysis concludes that ViSQOL\textsubscript{aswb} and VMAF 4K outperform other audio and video quality metrics, respectively.

The proposed dataset and findings in this research open new possibilities for future studies on primarily, but not limited to, AV quality evaluation in 360 videos with ambisonic spatial audio. Furthermore, the dataset from experiments can be used to advance existing objective quality metrics as well as propose a new ones by employing ML/DL based models. Our future work is to extend subjective and objective analysis together with the development of an AV perceptual quality model for 360 content.

\squeezeup
\section{Acknowledgements} 
\squeezeup
We convey the acknowledgment to FORCE Technology, Bang \& Olufsen, Demant, GN Store Nord, Sonova, WSA and industrial partners who created the 360 audio-visual datasets under the HOA-SSR joint project, and to XRHub Team for the great help in dealing with technicalities, filming and field recording. We also thank VRTonung and Vtopia360 for providing additional high-quality AV datasets for testing. This research was cofounded by Danish ministry of science and tech and the Marie Skłodowska-Curie grant agreement No.765911 RealVision.
\squeezeup

\balance
\small{
\bibliographystyle{IEEEbib}
\bibliography{references}
}

\end{document}